\documentclass{article}
\usepackage{typearea,dcolumn,cite,amsmath}
\usepackage[dvips]{epsfig}
\newcommand{\mc}[1]{\multicolumn{1}{c}{#1}}
\newcommand{\mcl}[1]{\multicolumn{1}{c|}{#1}}
\begin{document}
\title{Constant Pressure Hybrid Molecular Dynamics$-$Monte Carlo Simulations}
\author{Roland Faller \and Juan J. de~Pablo\\
    \small Department of Chemical Engineering, University of Wisconsin,
    Madison, WI 53706}
\maketitle
\renewcommand{\thetable}{\Roman{table}}
\begin{abstract}
\noindent New hybrid Molecular Dynamics-Monte Carlo methods are proposed to 
increase the efficiency of constant-pressure simulations. Two variations of the
isobaric Molecular Dynamics component of the algorithms are considered. In the
first, we use the extended-ensemble method of Andersen~[H. C. Andersen J. Chem.
Phys. {\bf 72},2384 (1980)]. In the second, we arrive at a new
constant-pressure Monte Carlo technique based on the reversible generalization
of the {\it weak$-$coupling} barostat~[H. J. C. Berendsen {\it et. al} 
J. Chem. Phys. {\bf 81}, 3684(1984)]. This latter technique turns out to be 
highly  effective in equilibrating and maintaining a target pressure. It is 
superior to the extended-ensemble method, which in turn is superior to simple
volume-rescaling algorithms. The efficiency of the proposed methods is 
demonstrated by studying two systems. The first is a simple Lennard-Jones 
fluid. The second is a mixture of polyethylene chains of 200 monomers.
\end{abstract}
\section{Introduction}
Molecular simulations at constant pressure are attractive because they mimic 
the conditions often encountered in laboratory experiments. This applies to 
both Molecular Dynamics and Monte Carlo calculations. In Monte Carlo 
simulations, trial configurations are generated in a random manner, and 
accepted according to well-defined probability criteria. The so-called
``hybrid'' schemes constitute one of many ways of proposing such trial 
configurations. In a hybrid Monte Carlo trial move, a short molecular dynamics
trajectory is used to generate a new configuration~\cite{heermann90,mehlig92};
the move 
is subsequently accepted or rejected according to probability criteria which 
take into account the way in which the trial configuration was generated (since
molecules are moved in the direction of the forces acting on them, the 
resulting bias must be eliminated). Molecular Dynamics becomes exact in the 
limit of vanishing time$-$step; the acceptance rate of trial configurations 
produced by a short MD run with small time-steps can therefore be expected to 
be high. The use of longer time-steps does not deteriorate the accuracy of a 
hybrid scheme, but it decreases the acceptance rate. The Monte-Carlo
acceptance/rejection procedure ensures that the correct statistical-mechanical
ensemble is sampled, and the results remain exact, regardless of time-step. In 
contrast, a pure molecular dynamics simulation using large time-steps would 
lead to a fast algorithm but questionable results.

Many algorithms for isobaric-isothermal $NPT$ molecular dynamics have been 
developed over the past 
decades~\cite{andersen80,parinello82,berendsen84,martyna96}. In contrast, 
$NPT$ Monte Carlo simulations still rely on the use of simple, random volume
changes accompanied by uniform rescaling of the molecules' coordinates. A few 
exceptions have been reported in the literature, but they lack the ease of 
implementation of simple random volume moves~\cite{escobedo95}. Unfortunately,
volume relaxation is notoriously slow in simple, random volume-rescaling Monte 
Carlo algorithms. Furthermore, their efficiency deteriorates as the size of a 
system increases.

Our experience indicates that volume moves are actually one of the limiting 
steps towards successful Monte Carlo simulations of complex fluids. It has, 
for example, become clear in studies of solubilities of small molecules in 
polymers that the computational bottleneck is indeed the equilibration of the 
density~\cite{nath00}. Interestingly, literature results suggest that
constant-pressure molecular dynamics algorithms can provide relatively fast 
volume relaxation. It is therefore of interest to explore whether a hybrid 
$NPT$ method is able to circumvent some of the problems associated with 
conventional $NPT$ Monte Carlo techniques. This paper considers two hybrid 
$NPT$ approaches. We find both of these to be superior to conventional methods
for constant-pressure Monte Carlo simulations.
\section{Constant Pressure Monte Carlo}
Before presenting the two hybrid methods to be discussed here, we shortly
review the fundamentals of Monte Carlo simulations in the
isothermal$-$isobaric ensemble. The natural variables are particle number $N$,
pressure $P$, and temperature $T$; the statistical mechanical potential for
this set of variables is the Gibbs free energy $G(N,P,T)$. As $N$, $P$ and $T$
are fixed, their conjugate variables must fluctuate. These are the chemical
potential $\mu$ (which, for a pure system, is equal to the Gibbs free energy
density $G/N$), the system volume $V$ and the entropy $S$.

The relevant probability distribution function for an $NPT$ ensemble is given
by
\begin{equation}
  f_{NPT} = \exp[-\beta U(\{\vec{r}\},\vec{p}\})-\beta P^* V(\{\vec{r}\})].
  \label{eq:basis}
\end{equation}
Here $P^*$ is the system pressure, $\{\vec{p}\}$ is the set of all momenta, 
$\{\vec{r}\}$ the set of all positions, $U$ the internal energy of the system 
and $V$ its volume; $\beta=1/(k_BT)$, where $k_B$ is Boltzmann's constant and 
$T$ is the temperature.

This distribution function is to be sampled by a simulation in an $NPT$ 
ensemble; the following acceptance criteria for trial moves generate
configurations distributed according to Equation (\ref{eq:basis}):
\begin{equation}
  p_{acc} =\min\{1,\exp[-\beta \Delta U(\{\vec{r},\{\vec{p}\})
  -\beta P^* \Delta V(\{\vec{r}\})]\}, \label{eq:metropolis}
\end{equation}
where $\Delta$ denotes the difference of a property between the new (proposed) 
state and the old (original) state, and where $\min(x,y)$ denotes the minimum 
of its two arguments. The condition of detailed balance, namely
\begin{eqnarray}
  K(a \to b) &=& K (b \to a)\nonumber\\
  K(a\to b) &=& p(a)q(a\to b)p_{acc}(a \to b)\label{eq:rate},
\end{eqnarray}
facilitates the construction of a correct algorithm. In 
Equation~(\ref{eq:rate}), $K(a \to b)$ is the probability of going from state 
$a$ to state $b$; this probability can be written as the product of $p(a)$, the
probability being in state $a$, $q(a\to b)$, the probability of proposing a 
transition from $a$ to $b$, and $p_{acc}(a \to b)$, the probability of 
accepting such a transition. Accepting or rejecting trial moves according to
\begin{equation}
  p_{acc}(a \to b) = \min \left(1,\frac{p(b) q(b\to a)}{p(a) q(a\to b)}\right) 
\label{eq:acceptance}
\end{equation}
ensures that configurations are distributed according to the desired 
probability distribution $p$ appearing in Equation~(\ref{eq:rate}).
\section{The extended ensemble algorithm}
In order for a Molecular Dynamics algorithm to be suitable for direct
application within a hybrid Monte Carlo formalism, two conditions must be
fulfilled. The algorithm should be reversible in time and symplectic, i.e.
preserving phase$-$space volume. If these requirements are met,
Equation~(\ref{eq:rate}) is fulfilled and the correct Markov process is
produced. Given these constraints, the Verlet integration scheme has often been
used in this context~\cite{verlet67}. It is reversible, it preserves
phase$-$space volume, and it is correct to order $O(\Delta t^3)$.

The extended-ensemble technique of Andersen~\cite{andersen80} (in conjunction
with the Verlet integrator) satisfies the above prerequisites. Its application
in the context of a hybrid move is therefore promising. The relevant equations
of motion are given by (for a derivation see the original paper of
Andersen~\cite{andersen80}):
\begin{eqnarray}
  \text{d}_t\vec{r}_i&=&\frac{p_i}{m_i}+\frac{1}{3}\text{d}_t \ln V\nonumber\\
  \text{d}_t\vec{p}_i&=&-\sum_{j\prime=1}^{N}\hat{r}_{ij}u^{\prime}(r_{ij})
  \nonumber\\
  M\text{d}^2_tV&=&-\alpha+\frac{1}{V}\left(\frac{2}{3}\sum_{i=1}^{N}
  \frac{p_i^2}{2m}-\frac{1}{3}\sum_{i<j=1}^Nr_{ij}u^{\prime}(r_{ij})\right),
  \label{eq:andersenset}
\end{eqnarray}
where $M$ is a fictitious mass of a ``piston'' associated with the volume move
(the piston connects the system to an infinitely large pressure reservoir).
Parameter $\alpha$ describes the fictitious potential of the volume. These
equations can be easily implemented, and the resulting molecular dynamics
trajectory obeys a constant $NPH$ constraint. Anderson's method can be
generalized to constant $NPT$, but this is not of interest here as maintaining
a constant temperature is left to the Monte Carlo acceptance criteria.

In most Monte Carlo simulations one deals with particle positions only. 
Therefore, at the beginning of a hybrid step, velocities (or momenta) must be 
assigned to each particle. These are generated at random from a 
Maxwell-Boltzmann distribution. In addition, a ``volume'' velocity $v_V$ is 
also generated.

The acceptance criteria for a volume hybrid move are derived from
Equations~(\ref{eq:rate}) and (\ref{eq:acceptance}). Since the integration of 
the equations of motion itself is deterministic, after having assigned 
velocities to the molecules, the probability of reaching a final state from 
some original state is unity. Still, the probability of assigning velocities 
to the particles must be considered, thereby leading to a transition 
probabilities of the form:
\begin{eqnarray}
  q( a\to b) & = & E_{kin}^{(\text{a})} p(v_V^{(\text{a})}) \nonumber \\
  q( b\to a) & = & E_{kin}^{(\text{b})} p(v_V^{(\text{b})})
  \label{eq:transition}
\end{eqnarray}
%
%
where $p(v_V)$ is the probability of generating a volume velocity according to 
a specific distribution, $E_{kin}$ is the kinetic energy, and superscripts $a$ 
and $b$ serve to denote the original and new (trial) states, respectively. The
acceptance criteria for hybrid $NPT$ moves are obtained by substituting 
Equations~(\ref{eq:transition}) into Equation~(\ref{eq:acceptance}).
\section{The reversible Weak-Coupling algorithm}
One of the most efficient algorithms for constant temperature and/or pressure 
Molecular Dynamics is the {\it weak-coupling} scheme proposed by Berendsen 
{\it et al.}~\cite{berendsen84}. Although it is widely used, unlike Andersen's
algorithm, it is not time-reversible due to the first-order nature of the 
equations of motion. This precludes its direct use in hybrid Monte$-$Carlo 
moves. The algorithm proposed in what follows eliminates this shortcoming.

Recently, it has been shown that the Weak-Coupling {\it thermostat} produces a
correct ensemble, in agreement with other thermostat methods for any value of
the coupling parameter $\tau$~\cite{morishita00}. We briefly describe the
Weak-Coupling algorithm, focusing only on the constant pressure case. In any
Molecular Dynamics algorithm the integrator moves the velocities and positions
in a time-step according to the equations of motion:
\begin{eqnarray}
  \vec{r}_i(t) \to \vec{r}_i(t+\Delta t)\\
  \vec{v}_i(t) \to \vec{v}_i(t+\Delta t)
\end{eqnarray}
After this update, the weak-coupling algorithm calculates the new pressure $P$
of the system based on the positions and velocities. This pressure is compared
to the desired, target pressure $P^*$. The volume of the simulation box is now
scaled ``towards'' the target pressure, i. e.
\begin{equation}
  V(t+\Delta t) = \tau_p^{-1}V(t) \cdot \Big(\frac{P(t)- P^*}{P^*}\Big)\Delta t
  \label{eq:berend}
\end{equation}
where $\tau_p$ is a characteristic time for this relaxation process (see 
below), and $\Delta t$ is usually (although not necessarily) the same 
time-step used to integrate the equations of motion. The positions of all 
particles must be adapted to the new volume, i.e. rescaled by a factor 
$V^{1/3}$. For convenience and efficiency, the molecular dynamics step is 
performed in scaled coordinates ranging from $s=\frac{r}{V^{1/3}}=[-0.5,0.5]$.

The smaller the relaxation $\tau_p$, the more closely the instantaneous 
pressure is tied to the target, and the stronger the disturbance of the actual
dynamics by individual rescaling operations. In molecular dynamics, $\tau_p$ 
must be large enough to produce meaningful data, as the fluctuations of 
pressure produced by this algorithm are not entirely correct~\cite{paci96}.
This does not pose a problem for Monte Carlo, as the molecular dynamics moves 
are only used to propose trial configurations.

As pointed out earlier, the volume rescaling of the Weak-Coupling algorithm is
not time-reversible. This violates the {\it detailed balance} rate
equation~(\ref{eq:rate}). The solution to this problem is to allow the
algorithm to run explicitly backward and forward in time; at the beginning of
a hybrid move, a random number $\zeta$ is drawn between 0 and 1. If
$\zeta<0.5$, the time direction in the simulation runs forward; otherwise we
run backwards. Running the simulation in reverse is equivalent to setting the
relaxation time $\tau_p$ to a negative value, as the rest of the equations of
motion is symmetric under time-reversal. The acceptance criteria for this
constant pressure hybrid move is similar to that for the Andersen case.

The weak-coupling method as outlined so far results in rather slow
fluctuations around the desired pressure. Furthermore, (if backward moves are
momentarily disregarded) the relaxation becomes exponentially slow close to
the target pressure. This limits the efficiency of the simulation. To improve
efficiency, we generalize the weak-coupling hybrid scheme to a target pressure
selected randomly within a bounded domain. The pressure $P^*$ used as {\it
local} target pressure for the simulation in one hybrid step is now chosen
with uniform probability from the interval
$[(1-P_{\text{range}})P^{(\text{ext})},(1+P_{\text{range}})P^{(\text{ext})}]$,
whereas the pressure in the acceptance criteria remains untouched
(at $P^{(\text{ext})}$). Values for $P_{\text{range}}$ range from 0 to 0.1.
Since the {\it local} target pressure is uniformly chosen from the neighborhood
of the {\it global} target pressure, detailed balance is still obeyed.
\section{Efficiency Considerations}
The main parameter which can be tuned and optimized in the reversible
weak-coupling (RWC) simulation is the relaxation time $\tau_p$. The smaller
$\tau_p$, the closer the system follows the desired target pressure, but the
lower the acceptance rate is.

\begin{figure}
  \includegraphics[angle=-90,width=0.5\linewidth]{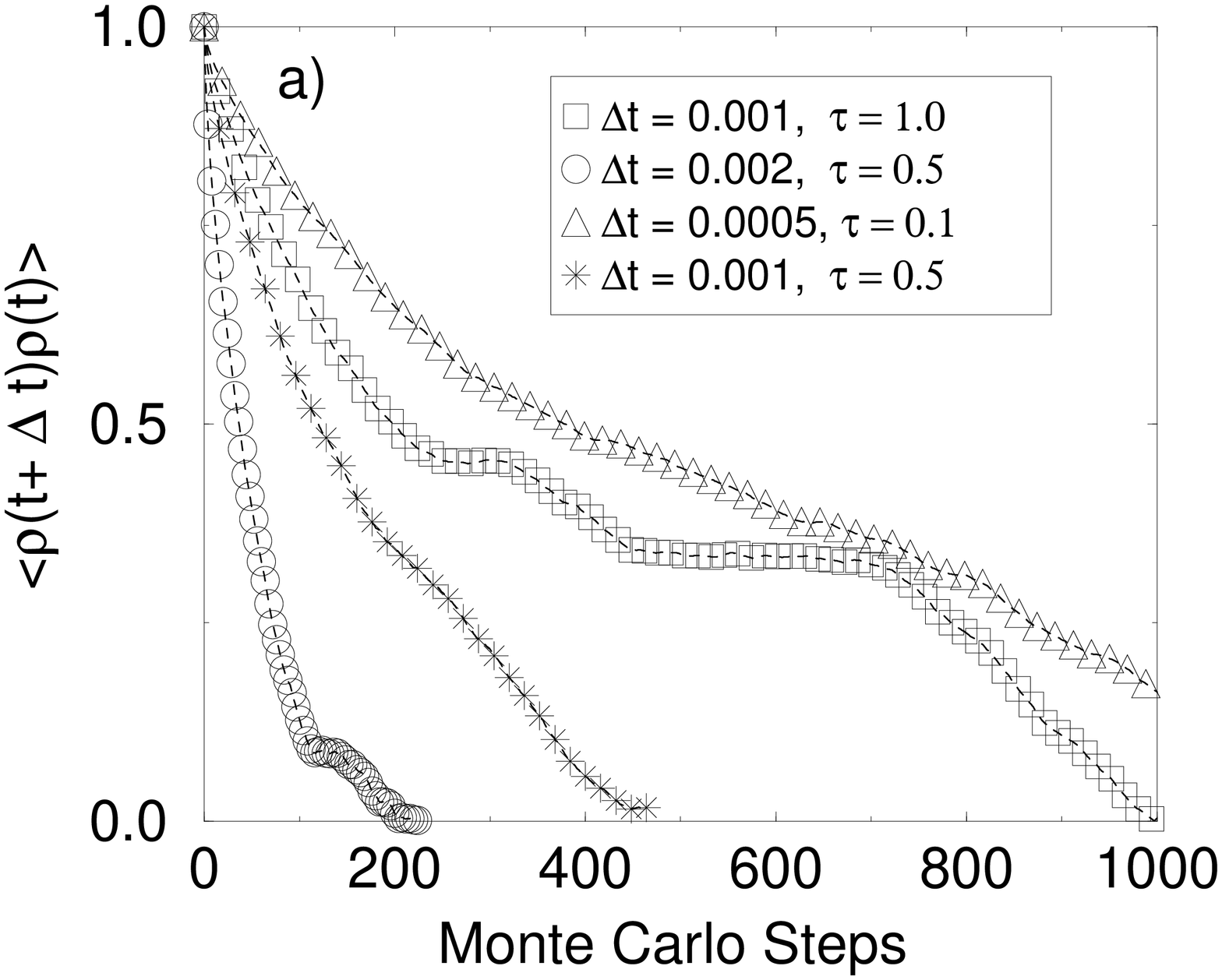}
  \includegraphics[angle=-90,width=0.5\linewidth]{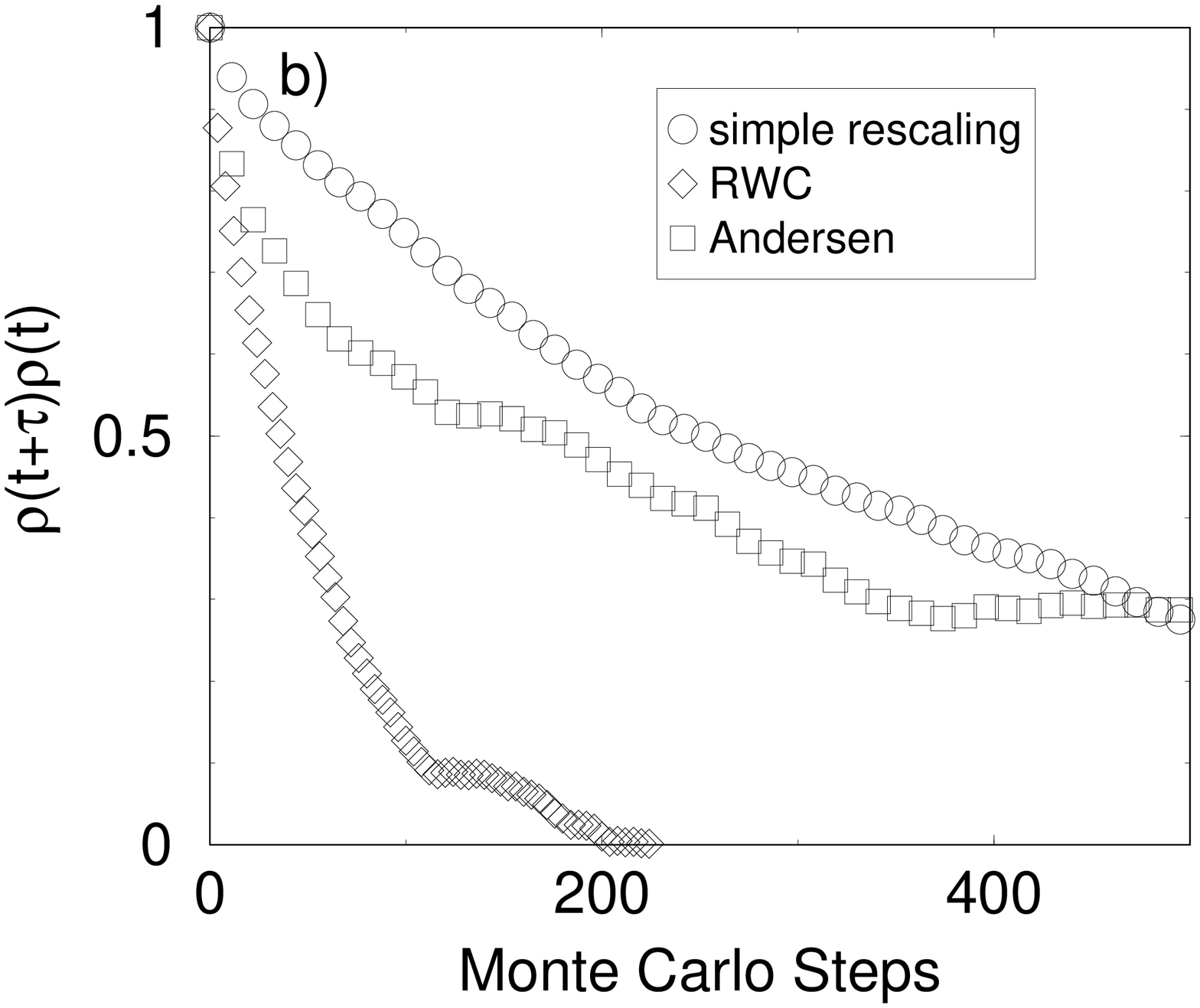}
  \caption{Density Autocorrelation function for a Lennard-Jones fluid
  ($T=1.4,\,p=0.21732, \to\rho=0.45$) with different values of time-step and
  correlation time (part a). $p_{\text{range}}=0.1$ the values for the
  timestep and the correlation time are given in the graph. Part b) shows the
  comparison to simple rescaling and the hybrid method based on the Andersen
  barostat. The box mass $M$ in the Andersen case was set to $M=1.0$.}
\label{fig:effiber}
\end{figure}

We analyze the density autocorrelation function to provide a measure of
efficiency for the proposed hybrid schemes. Figure~\ref{fig:effiber}a) shows
that, for a simple Lennard Jones fluid, increasing the time-step leads to
smaller correlation times and faster equilibration of the density. However, if
the ratio of correlation time to time step becomes too small the simulation
becomes again inefficient as one would need extremely large neighbor
lists (or extremely frequent updates) to balance the rapid motion. The figure
shows how important a good choice of the parameters can be. An unfortunate
choice results in only marginal improvements relative to conventional methods.

The ratio between the time-step and the correlation time is important. The
correlation function for $\Delta t=0.001, \tau=0.25$ is the same as for
$\Delta t=0.002, \tau=0.5$ (not shown). We found that a value of this ratio of
about 250 provides good efficiencies. If we increase this ratio the acceptance
rate drops; if we decrease it the efficiency of the single move deteriorates.

The conventional method for constant pressure Monte-Carlo simulations
comprises simple volume-rescaling trial moves. In such moves a random change
of volume is proposed according to
\begin{equation}
  V_{\text{new}}=V_{\text{old}}+\Delta V,
\end{equation}
with $\Delta V$ chosen uniformly from
$[-\Delta V_{\text{max}},\Delta V_{\text{max}}]$.

Our results indicate that, even for a simple Lennard-Jones fluid, the hybrid
techniques proposed here provide methods that appear to be superior to the
simple volume-rescaling technique. For a good choice of parameters, the
efficiency gains can amount to more than an order of magnitude.
\section{Test cases}
\subsection{The Lennard-Jones Fluid}
As a test case, the Lennard-Jones fluid is revisited. We simulate a system of
200 Lennard-Jones particles at a range of temperature and pressures. The
cutoff is $r=2.5\sigma$ and the usual analytic long-range corrections for
energy and pressure are applied~\cite{frenkel96}. All values are given in the
standard dimensionless units. The simulations used 1 million MC steps, each
containing 5 constant-pressure molecular dynamics integration steps.

We compare our results to those obtained from an equation of
state~\cite{johnson93} which is known to provide accurate properties for the
Lennard-Jones fluid. At a given temperature and pressure we compare density
and internal energy. The results are shown in table~\ref{tab:cmpLJ}; agreement
between simulations and equation of state predictions is excellent, showing
that the algorithms proposed here lead to correct property predictions. Note
that we simulated the same systems (using identical starting conditions) using
the three methods considered in this work (Andersen, Weak-Coupling, and Simple
Volume-Rescaling). There were no noticeable differences in the results.

\begin{table}
  \begin{center}
  \begin{tabular}{D{.}{.}{-1}D{.}{.}{-1}|D{!}{.}{-1}r|D{.}{.}{-1}D{.}{.}{-1}}
  \hline
  \mc{T}& \mcl{p}& \mc{$\rho_1$} &\mcl{$u_1$}& \mc{$\rho_2$}& \mc{$u_2$} \\
  \hline
  0.75 & 1.59 & 0!90\pm0.01   & $-6.35\pm0.03$ & 0.904 & -6.40 \\
  1.6  & 0.1231 & 0!096\pm0.002 & $-0.73\pm0.03$ & 0.094& -0.67\\
  1.6  & 0.16 & 0!132\pm0.001 & $-0.91\pm0.02$ & 0.132 & -0.91 \\
  1.6  & 0.38 & 0!42\pm0.01   & $-2.79\pm0.03$ & 0.408 & -2.72 \\
  1.6  & 5.07 & 0!84\pm0.01   & $-5.10\pm0.03$ & 0.85  & -5.27 \\
  \hline
  \end{tabular}
  \caption{Comparison of density and internal energy per particle for the
  Lennard-Jones system from simulation of this work using reversible weak
  coupling (1) and from the work of Johnson {\it et al.} (2).}
  \label{tab:cmpLJ}
  \end{center}
\end{table}

We now examine the volume fluctuations produced by the different methods. 
These fluctuations are of interest because they provide a stringent test of the
correctness and the efficiency of a simulation technique, and they are related
to the compressibility of the fluid. We compare our results to the equation of
state by Rosenfeld~\cite{rosenfeld98}, again at $T=1.6$ (cf.
Table~\ref{tab:compress}). As expected, the compressibility corresponding to 
our various simulation techniques agrees with accepted values.
\begin{equation}
  \kappa_T=\frac{\langle V^2\rangle-\langle V\rangle^2}{k_BTV}.
\end{equation}

\begin{table}
  \begin{center}
    \begin{tabular}{D{.}{.}{-1}D{.}{.}{-1}D{.}{.}{-1}}
      \hline
      \mc{$\rho$} & \mc{$\kappa_{T1}$} & \mc{$\kappa_{T2}$} \\
      \hline
      0.13 & 6.7\pm2     & -- \\
      0.40 & 2.4\pm0.5   & 2 \\
      0.84 & 0.04\pm0.01 & 0.05\\
   \end{tabular}
   \caption{Isothermal compressibilities for the Lennard-Jones fluid at $T=1.6$
   calculated by fluctuations of the volume ($\kappa_{T1}$) and by
   the equation of state by Rosenfeld ($\kappa_{T2}$). }
   \label{tab:compress}
\end{center}
\end{table}

\subsection{Polymer Melts}
In order to examine the performance of the proposed hybrid methods in the
context of a complex fluid, we used them to equilibrate a polymer melt.

Systems consisting of linear polyethylene chains (C$_{100}$ to C$_{500}$) and 
a small fraction of ethylene were placed in a simulation box at a density above
the experimental value. Three identical systems were left to equilibrate by the
reversible weak-coupling hybrid methods and the volume-rescaling algorithm.
The weak-coupling algorithm proved to be efficient and produced reliable, 
well-equilibrated densities for the pressure range under study ($200-500kPa$). 
All studies have been performed using the NERD 
force$-$field~\cite{nath00,nath01}.

\begin{figure}
  \begin{center}
  \includegraphics[angle=-90,width=0.45\linewidth]{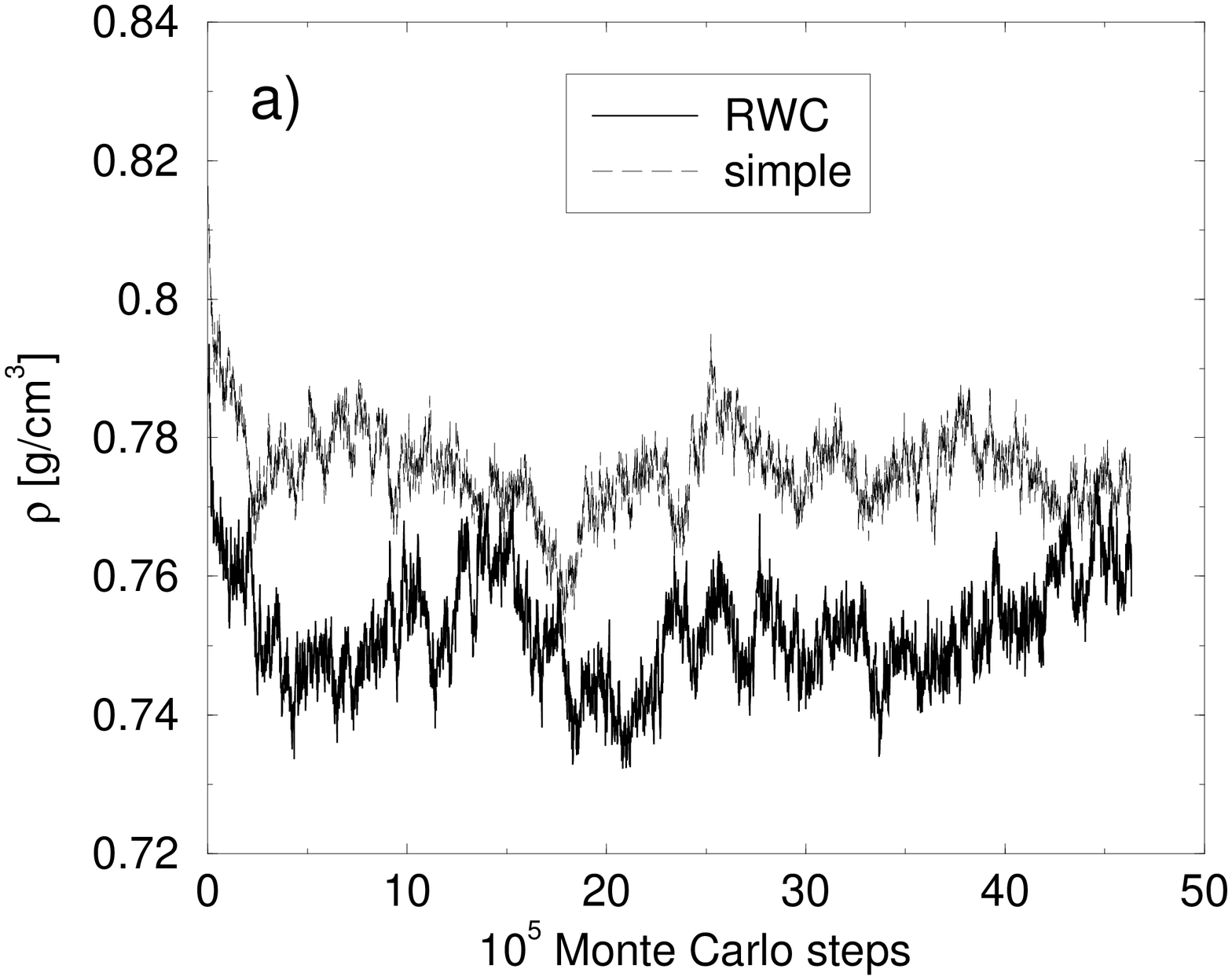}
  \includegraphics[angle=-90,width=0.45\linewidth]{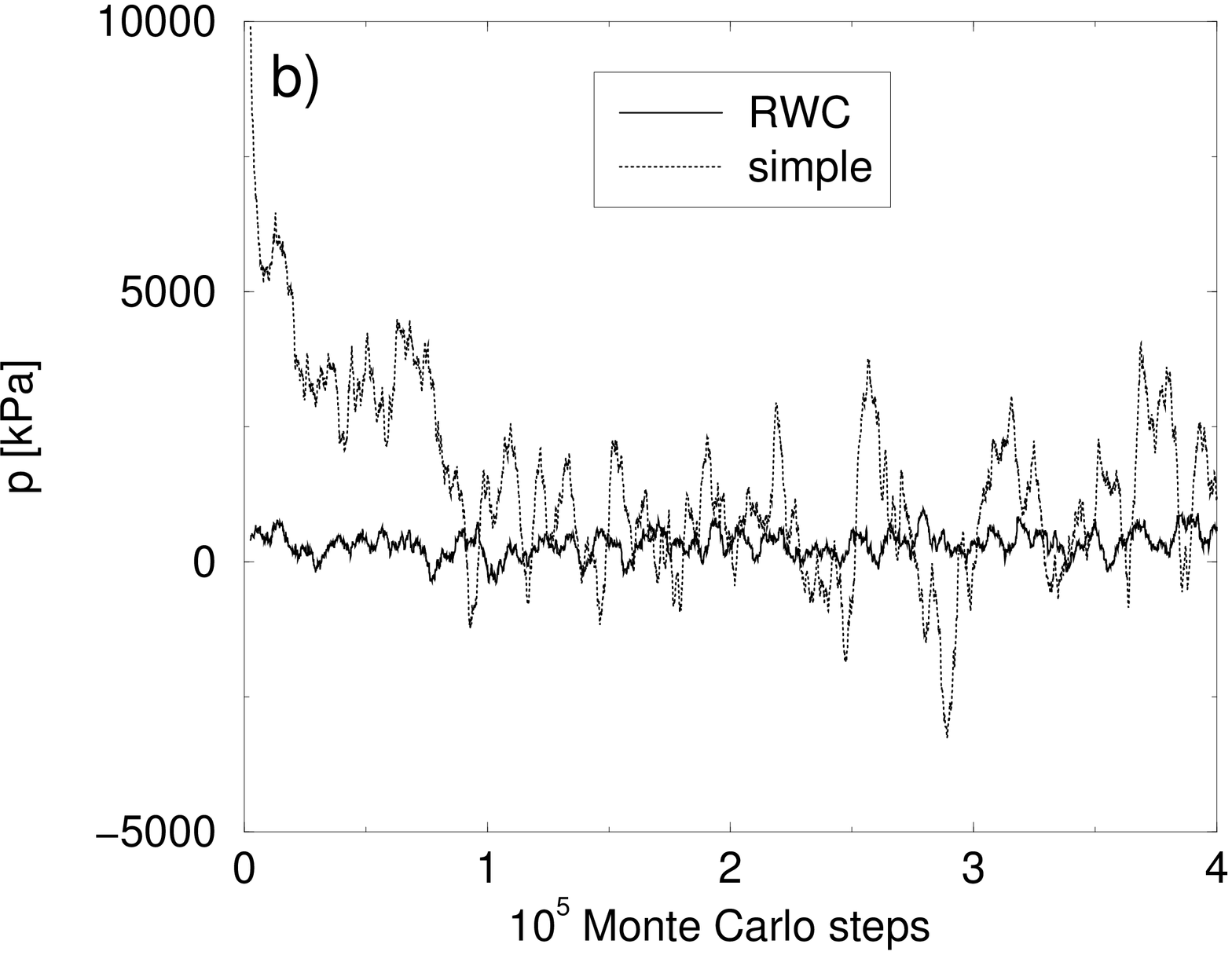}
  \caption{a) Density and b) Pressure as a function of simulation time for
  the reversible Weak-Coupling algorithm an the volume-rescaling algorithm for
  a system of 11 alkane chains of length 200 and 26 ethylene molecules at
  $P=260kPa,\,T=400K$. For the pressure, a running average over 50 data points
  clarity.}
  \label{fig:eff1}
  \end{center}
\end{figure}

Figure~\ref{fig:eff1} shows a comparison between the convergence of the newly 
proposed weak-coupling method and simple volume rescaling. Both simulations 
started at a system density of $0.82\; g/cm^3$, which is slightly too high. The
equilibrium density is around $0.75 \;g/cm^3$. It can be seen in the figure 
that the convergence of the weak-coupling algorithm to the correct, target
pressure is relatively fast. In contrast, pressure (or volume) relaxation in t
he conventional volume-rescaling simulation is slow. Even after $500,000$ 
steps, the running-average pressure in the conventional volume-rescaling 
simulation is far from the correct, target result (by about a factor of 3).

\section{Conclusions}
New hybrid schemes have been presented for $NPT$ Monte Carlo simulations. We 
find that the proposed methods are more effective at relaxing the volume of 
simple and complex liquids than conventional Monte Carlo techniques. The 
increase of efficiency becomes more apparent in fluids that exhibit long 
relaxation times (e.g. polymers).

An additional advantage of the methodologies discussed in this work is that 
available simulation codes for molecular dynamics calculations can be modified 
to perform exact Monte Carlo calculations with minimal changes. Implementation
of the proposed techniques does not add a significant overhead, and the new 
moves not only change the volume, but also contribute significantly to the 
repositioning of particles, thereby improving sampling.

It is our experience that equilibrating the density of a dense polymer melt 
using conventional $NPT$ Monte Carlo moves is highly computationally 
demanding, and often not possible at all. In fact, we have shown in this work 
that, as a result of poor sampling, the compressibility of a melt of short 
polyethylene chains determined using simple rescaling techniques deviates 
considerably from the correct value. In contrast, the compressibility 
predicted using the weak-coupling algorithm is in agreement with experimental
values.
\section*{Acknowledgements}
Discussions with Brian Banaszak and Kevin van Workum are gratefully
acknowledged. RF thanks the DFG (German Research Foundation) for financial
support in the framework of the Emmy-Noether program.
\bibliography{standard}

\begin{thebibliography}{10}

\bibitem{heermann90}
D.~W. Heermann, P.~Nielaba, and M.~Rovere,
\newblock Comp. Phys. Commun. {\bf 60}, 311 (1990).

\bibitem{mehlig92}
B.~Mehlig, D.~W. Heermann, and B.~M. Forrest,
\newblock Phys. Rev. B {\bf 45}, 679 (1992).

\bibitem{andersen80}
H.~C. Andersen,
\newblock J. Chem. Phys. {\bf 72}, 2384 (1980).

\bibitem{parinello82}
M.~Parinello and A.~Rahman,
\newblock J. Chem. Phys. {\bf 76}, 2662 (1982).

\bibitem{berendsen84}
H.~J.~C. Berendsen, J.~P.~M. Postma, W.~F. van Gunsteren, A.~DiNola, and J.~R.
  Haak,
\newblock J. Chem. Phys. {\bf 81}, 3684 (1984).

\bibitem{martyna96}
G.~J. Martyna, M.~E. Tuckerman, D.~J. Tobias, and M.~L. Klein,
\newblock Molecular Physics {\bf 87}, 1117 (1996).

\bibitem{escobedo95}
F.~A. Escobedo and J.~J. {de Pablo},
\newblock Macromolecular Theory and Simulations {\bf 4}, 691 (1995).

\bibitem{nath00}
S.~K. Nath and J.~J. {de~Pablo},
\newblock Mol. Phys. {\bf 98}, 231 (2000).

\bibitem{verlet67}
L.~Verlet,
\newblock Phys. Rev. {\bf 159}, 98 (1967).

\bibitem{morishita00}
T.~Morishita,
\newblock J. Chem. Phys. {\bf 113}, 2976 (2000).

\bibitem{paci96}
E.~Paci and M.~Marchi,
\newblock J. Phys. Chem. {\bf 100}, 4314 (1996).

\bibitem{frenkel96}
D.~Frenkel and B.~Smit,
\newblock {\em Understanding Molecular Simulation: From Basic Algorithms to
  Applications},
\newblock Academic Press, San Diego, CA, 1996.

\bibitem{johnson93}
J.~K. Johnson, J.~A. Zollweg, and K.~E. Gubbins,
\newblock Molecular Physics {\bf 78}, 591 (1993).

\bibitem{rosenfeld98}
Y.~Rosenfeld,
\newblock Molecular Physics {\bf 94}, 929 (1998).

\bibitem{nath01}
S.~K. Nath, B.~J. Banaszak, and J.~J. {de~Pablo},
\newblock J. Chem. Phys. {\bf 114}, 3612 (2001).

\end{thebibliography}
\bibliographystyle{aip}

\end{document}